\documentclass[epj]{svjour}
\usepackage{graphicx,times}
\usepackage{amssymb}

\newcommand*{\ua}{\uparrow}
\newcommand*{\da}{\downarrow}
\newcommand*{\ep}{\epsilon}
\newcommand*{\abs}[1]{\left|#1\right|}
\newcommand*{\absa}[1]{| #1 |}
\newcommand*{\aver}[1]{\left<#1\right>}
\newcommand*{\averaa}[1]{\langle #1\rangle}
\newcommand*{\bra}[1]{\left<#1\right|}
\newcommand*{\ket}[1]{\left|#1\right>}
\newcommand*{\braaa}[1]{\langle#1|}
\newcommand*{\ketaa}[1]{|#1\rangle}

\begin{document}
\title{Entanglement and transport through correlated quantum dot}

\author{Adam Rycerz}
\institute{
   Instituut--Lorentz, Universiteit Leiden,
   P.O.\ Box 9506, NL--2300 RA Leiden, The Netherlands \\
   Marian Smoluchowski Institute of Physics, Jagiellonian University, 
   Reymonta 4, 30-059~Krak\'{o}w, Poland\\
   \email{adamr@th.if.uj.edu.pl}}
\date{\today}
\abstract{
We study quantum entanglement in a single--level quantum dot in the
linear--response regime. 
The results show, that the maximal quantum value of the conductance 
$2e^2/h$ not always match the maximal entanglement. 
The pairwise entanglement between the quantum dot and the nearest atom of the 
lead is also analyzed by utilizing the Wootters formula
for \emph{charge} and \emph{spin} degrees of freedom separately.
The coexistence of zero concurrence and the maximal conductance is observed
for low values of the dot--lead hybridization.
Moreover, the pairwise concurrence vanish simultaneously for charge and
spin degrees of freedom, when the Kondo resonance is present in the system.
The values of a Kondo temperature, corresponding to the zero--concurrence
boundary, are also provided.
}
\authorrunning{A.\ Rycerz}
\titlerunning{Entanglement and transport through correlated quantum dot}
\maketitle


\section{Introduction}
Quantum entanglement, as one of the most intriguing features of quantum
mechanics, was extensively studied during the last decade, 
mainly because its nonlocal connotation \cite{epr} is regarded as a valuable 
resource in quantum communication and information processing \cite{qinfo}.
The question about the relation between the entanglement and quantum phase 
transitions \cite{qptra} have been addressed recently, for either quantum 
spin \cite{spinc,delga,oles,webza} and fermionic \cite{schli,zanar,guet} 
systems, in hope to shed new lights on fundamental problems of condensed 
matter physics. 
For example, it was shown for spin model \cite{spinc}, that the entanglement
of two neighboring sites displays a sharp peak either near or at critical
point where quantum phase transition undergoes.
Recently, a class of systems
with divergent entanglement length away from quantum critical point, since
the correlation length remains finite, was identified \cite{delga}.
The spin--orbital entanglement analysis was also shown to provide a valuable 
insight to the nature of Mott insulators \cite{oles}.
In the field of fermionic systems,
the local entanglement was successfully used to identify quantum
phase transitions in the extended Hubbard model \cite{guet}.
A separate issue concerns using the entanglement as a criterion of
quantum coherence \cite{coher} when analyzing nonequilibrium dynamics of the 
system with spontaneous symmetry breaking \cite{webza}.

Here we follow the above ideas, but focus on the physical system which 
undergoes the crossover behavior instead of a phase transition: a quantum dot 
in the Kondo regime.
\emph{Namely}, we address the question 
\textit{whether there exist a relation between entanglement and conductance} 
for this system?
Some earlier study mentioned the total entanglement of electronic
degrees of freedom in the $SU(4)$ system below the Kondo temperature, without 
determining a~qualitative measure of such an entanglement \cite{choi}.
In this paper, we consider the $SU(2)$ case, and analyze two 
different definitions of the entanglement between quantum dot and the leads:
\emph{first} based on the von Neumann entropy, and \emph{second} utilizing the 
Wootters formula \cite{woot} for the formation concurrence of two--qubit 
system.


\section{The model and its numerical solutions}
\begin{figure}[!t]
\begin{center}
\includegraphics[width=0.8\columnwidth]{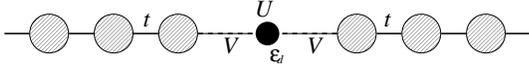}
\end{center}
\caption{
    The Anderson impurity model realized as a double quantum dot
    attached to the leads. The dot is described with the energy level 
    $\epsilon_d$ and the Coulomb interaction $U$.
}
\label{qdfig}
\end{figure}

We study a model of a quantum dot with a single relevant
electronic level coupled to the left $(L)$ and right $(R)$ metallic 
electrodes. The Hamiltonian of the system is
\begin{equation}
\label{ham5}
  H = H_L+V_L+H_C+V_R+H_R,
\end{equation}
where $H_C$ models the central region, 
$H_{L(R)}$ describes the left (right) lead itself, and $V_{L(R)}$ is the 
coupling between the lead and the central region. Namely, we have
$$
  H_C = \ep_d n_d + U n_{d\ua}n_{d\da},
$$
\begin{equation}
  H_{L(R)}=-t\sum_{\stackrel{j,j+1\in L(R)}{\sigma=\ua,\da}}
  \left( c^{\dagger}_{j\sigma}c_{j+1,\sigma} + \mathrm{h.c.} \right),
\end{equation}
$$
  V_{L(R)}=-V\sum_{\sigma}
  \left( c^{\dagger}_{j_{L(R)}\sigma}d + \mathrm{h.c.} \right).
$$
Here, $n_d=\sum_{\sigma}d^\dagger_\sigma d_\sigma$ is the quantum--dot charge, 
$\ep_d$ is the position of the molecular level and $U$ is the Coulomb 
repulsion between two electrons.
Both $H_{L(R)}$ and $V_{L(R)}$ terms have a tight--binding form, with the 
hopping $t$ and the dot--lead hybridization parameter $V$, 
$c^{\dagger}_{j\sigma}$ ($c_{j\sigma}$) creates (destroys) an electron with
spin $\sigma$ on site $j$, the indexes $j_{L(R)}$ denotes terminal sites
of the left (right) electrode.
The system is depicted schematically in~Fig.\ \ref{qdfig}.

There are many theoretical methods in the existing literature, developed to 
study the electron transport in the presence of interaction. 
In particular, the zero--temperature
conductance of the quantum dot acting as an Anderson impurity were obtained
within the \emph{Bethe ansatz} approach \cite{bean}. For the more general
situation, one can refer to the \emph{Numerical Renormalization Group} 
\cite{hofs} or to the nonequilibrium \emph{Keldysh formalism} \cite{meir}.
For example, the former approach was succesfuly generalized to study a
molecule with the electron--phonon coupling \cite{corn}, whereas
the latter was adapted for an analysis of the competition between the Fano and 
the Kondo resonance in various nanodevices \cite{bulk}.

However, since we are interested \emph{either} in equilibrium transport 
properties \emph{or} in the ground--state quantum entanglement, the most useful
choice is the variational method 
recently proposed by Rejec and Ram\v{s}ak \cite{rera,reram}, in which the 
real--space correlation functions are obtained directly. 
For the system described by the Hamiltonian (\ref{ham5}) the method converges 
to the exact  solution \cite{bean}, it can also be generalized for multiple 
quantum dots \cite{rera}, for the case with a nonzero magnetic field 
\cite{reram}, or combined with an \emph{ab initio} wave--function readjustment
\cite{ryspa} in the framework of EDABI method \cite{edabi}.


\section{The quantum entanglement}
For the spin $s=1/2$ fermionic system, there are four possible local
states at each site, $\ketaa{\nu}_j=\ketaa{0}_j,\ketaa{\ua}_j,
\ketaa{\da}_j,\ketaa{\ua\da}_j$. The dimension of the $N$--site system
is then $4^N$ and $\ketaa{\nu_1,\nu_2,\dots,\nu_N}=
\prod_{j=1}^N\ketaa{\nu_j}_j$ are its natural basis vectors.
Alternatively, one can label the basis vectors by specifying occupation
numbers for each site and spin
$
  \ket{\nu_1 \dots \nu_N} \equiv 
  \ket{n_{1\ua} \dots n_{N\ua}}\ket{n_{1\da} \dots n_{N\da}}, 
$
with $n_{j\sigma} = 0,1$.
The reduced density matrix for the ground state $\ketaa{\Psi}$ is
\begin{equation}
\label{rhodef}
  \rho_{i\sigma,j\sigma'}=
  \mbox{Tr}_{i\sigma,j\sigma'}\ketaa{\Psi}\braaa{\Psi}, 
\end{equation}
where $\mbox{Tr}_{i\sigma,j\sigma'}$ stands for tracing over
all sites and spins except the $i\sigma$ and $j\sigma'$--th sites.

\subsection{Local entanglement and conductance}
We focus now on the local entanglement \cite{zanar}, which exhibits the 
quantum correlations between local state of a selected $j$--th site (e.g.\ 
the quantum dot) and the other part of the system (here: the leads). 
For $i\equiv j=d$ and $\sigma'\equiv\bar{\sigma}$, the reduced 
density matrix, defined by Eq.\ (\ref{rhodef}), takes the form 
\begin{equation}
\label{rhod}
  \rho_d = u_+\ket{0}\bra{0} + w_1\ket{\ua}\bra{\ua}
  + w_2\ket{\da}\bra{\da} + u_-\ket{\ua\da}\bra{\ua\da},
\end{equation}
where
$$
  u_+ = \aver{(1-n_{d\ua})(1-n_{d\da})},\ \ \ \ 
  w_1 = \aver{n_{d\ua}(1-n_{d\da})},
$$
\begin{equation}
  \label{delem}
  w_2 = \aver{(1-n_{d\ua})n_{d\da}},\ \ \ \ 
  u_- = \aver{n_{d\ua}n_{d\da}},
\end{equation}
and the averaging is performed for the system ground state.
Consequently, the corresponding von Neumann entropy 
$E_v$ (he\-re\-in\-af\-ter called the \emph{local entanglement}) measures the 
entanglement of the states of quantum dot ($j=d$) with that of the remaining 
$N-1$ sites, and is given by
\begin{equation}
\label{evdef}
  E_v = -u_+\log_2u_+ - w_1\log_2w_1
  - w_2\log_2w_2 -u_-\log_2u_-.
\end{equation}

\begin{figure}[!t]
\begin{center}
\includegraphics[width=0.8\columnwidth]{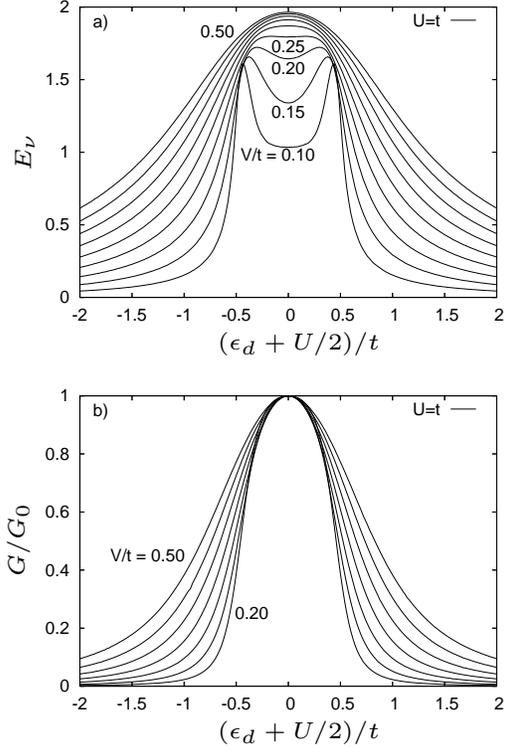}
\end{center}
\caption{
  The local entanglement $(a)$ and normalized conductance $(b)$ for the system 
  in Fig.\ \ref{qdfig} as a function of the dot energy level $\epsilon_d$ 
  and dot--lead hybridization $V$ (changed in steps of $0.05t$).
}
\label{gevep}
\end{figure}

In Fig.\ \ref{gevep} we compare the the local entanglement $E_v$ with the 
conductance calculated from the Rejec--Ram\v{s}ak \emph{two--point formula} 
\cite{rera}
\begin{equation}
\label{grera}
  G=G_0\sin^2\frac{\pi}{2}\frac{E(\pi)-E(0)}{\Delta},
\end{equation}
where $G_0=2e^2/h$ is the conductance quantum,
$\Delta=1/N\rho(\epsilon_F)$ is the average level spacing at Fermi energy,
determined by the density of states in an infinite lead $\rho(\epsilon_F)$,
$E(\pi)$ and $E(0)$ are the ground--state energies of the system with 
\emph{periodic} and \emph{antiperiodic} boundary conditions, respectively. 
We found that the system size of the order of $N\sim 1000$ 
provides an excellent convergence for both the conductance $G$
and the local entanglement $E_v$ (the latter aspect has not been analyzed 
numerically before). In particular, the data for $N=1000$ cannot be
distinguished from the ones for $N=2000$ in the scale of Fig.\ \ref{gevep}.
We also checked that the results are insensitive to the number of basis
functions composing the Rejec--Ram\v{s}ak variational wave--function 
\cite{rera}, providing it is $\geqslant 3$.
When calculating correlation functions (\ref{delem}), determining  the density 
matrix (\ref{rhod}), one has to choose boundary conditions which minimize the 
ground--state energy for a given system size $N$: namely, periodic for 
$N=4k+2$, and antiperiodic for $N=4k$ \cite{pabc} for the half--filling 
\cite{half}.

\begin{figure}[!t]
\begin{center}
\includegraphics[width=0.8\columnwidth]{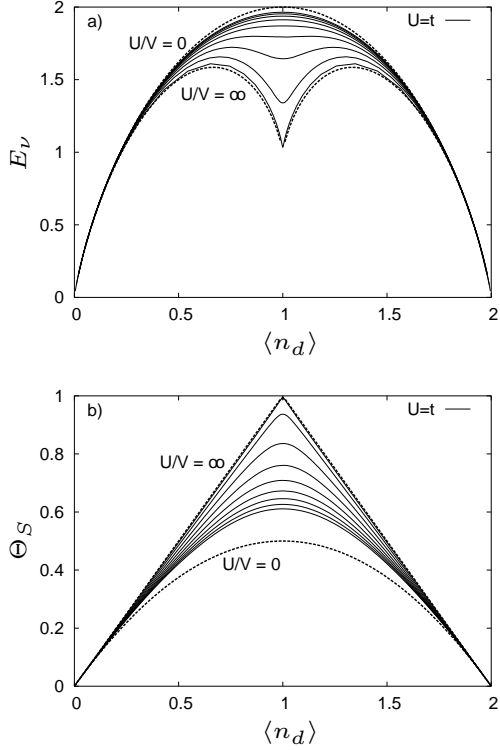}
\end{center}
\caption{
  The local entanglement $E_v$ $(a)$ and spin--magnitude parameter
  $\Theta_S\equiv(4/3)\averaa{\mathbf{S}_d^2}$ $(b)$ as a function of 
  the dot filling $\averaa{n_d}$ and dot--lead hybridization $V$.
  The magnitude of hybridization $V$ goes from $0.1t$ to
  $0.5t$ in steps of $0.05t$. The limiting curves for $U/V=0$ and 
  $U/V=\infty$ are depicted with \emph{dashed} lines.
}
\label{evs2nd}
\end{figure}

\emph{Surprisingly,} the maximal entanglement between a quantum dot and leads
not always match the maximal conductance $G=G_0=2e^2/h$. 
For small values of the dot--lead hybridization ($V\lesssim 0.25t$ for $U=t$), 
$E_v$ has a~minimum at the par\-ti\-cle--hole symmetric point 
$\epsilon_d=-U/2$. 
It is well known \cite{kondo} that in the limit $V^2/t\ll U$ the Anderson
Hamiltonian (\ref{ham5}) reduces to the symmetric Kondo model with an exchange
coupling $\rho(\epsilon_F)J_K=8V^2/\pi U\sqrt{4t^2-\epsilon_F}$.
This observation suggest an important role of the localized moment presence
in the dot, which strongly affects the entanglement between the quantum
dot and the leads, without observable change to the conductance.
The latter refers to the situation below the Kondo temperature
$T_K\propto \exp(-1/\rho(\epsilon_F)J_K)$, above which the conductance at the
particle--hole symmetric point $\epsilon_d=-U/2$ is depressed and, 
subsequently, each zero--$T$ Kondo peak displayed on Fig.\ \ref{gevep}b splits
into Coulomb--blockade peaks present in a finite--$T$ situation 
\cite{hofs,meir,corn,bulk}. 
In particular, for $U=t$ and $V=0.25t$, the exact formula 
\cite{tksym} gives the value of $T_K/t\approx 3\mathrm{mK/eV}$, which seems
to be in the experimentally accessible range.
A finite--$T$ analysis is, however, beyond the scope of this paper,
since we focus here on $T=0$ situation.
A further discussion of the relation between spin fluctuations and the
entanglement at the ground state is provided below.

For the better overview of the system properties we analyze them
as functions of the dot filling $\averaa{n_d}$, as displayed in
Fig.\ \ref{evs2nd}. 
The universal formula for the conductance (not shown) follows from the 
Luttinger theorem \cite{hews}
\begin{equation}
\label{glutt}
  G=G_0\sin^2(\pi\aver{n_d}/2),
\end{equation}
whereas $E_v$ evolves gradually from the limit 
\begin{equation}
\label{evzero}
  E_v^{U=0} = 
   - \averaa{n_d}\log_2\frac{\averaa{n_d}}{2}-\left(2\!-\!\averaa{n_d}\right)
     \log_2\left(1\!-\!\frac{\averaa{n_d}}{2}\right) 
\end{equation}
to
\begin{equation}
\label{evinfty}
  E_v^{U=\infty} =
   - \left(1\!-\!\averaa{n_d}\right)\log_2\left(1\!-\!\averaa{n_d}\right)
   - \averaa{n_d}\log_2\frac{\averaa{n_d}}{2},
\end{equation}
as presented in Fig.\ \ref{evs2nd}a.
Therefore, in the strong coupling limit $E_v$ has two maxima at 
$\averaa{n_d}=2/3$ and $4/3$, instead of a single one for $\averaa{n_d}=1$,
present in the noninteracting case. 
The major qualitative difference between our results and that obtained for
the extended Hubbard model \cite{guet} is that $E_v$ behaves analytically
for any $U<\infty$. This is because the Kondo system, considered here,
shows the crossover behavior instead of a quantum phase transition present
in the Hubbard chain.

One can also observe, that the 
entanglement be\-ha\-vior near the particle--hole symmetric point 
$\averaa{n_d}=1$ is determined by the magnitude of spin fluctuations,
presented in Fig.\ \ref{evs2nd}b.
As a measure of such fluctuations, we choose a parameter \cite{spal}
\begin{equation}
\label{thes}
  \Theta_S \equiv \frac{4}{3}\averaa{\mathbf{S}_d^2} =
  \aver{n_d} - 2\aver{n_{d\ua}n_{d\da}},
\end{equation}
which obeys the inequality 
\begin{equation}
\label{thlim}
  \averaa{n_d}\left(1-\frac{\averaa{n_d}}{2}\right)\leqslant \Theta_S 
  \leqslant 1-\abs{1-\averaa{n_d}},
\end{equation}
where the lower and the upper limit refers to the $U=0$
and $U=\infty$ case, respectively.
In particular, for $\averaa{n_d}=1$ the spin--fluctuation parameter varies 
from $\Theta_S=1/2$ for free fermions to $\Theta_S=1$ for the localized 
spin--$1/2$. 
The charge fluctuations are determined by $\Theta_S$ as 
$\mathrm{Var}\{n_d\}\equiv\averaa{n_d^2}-\averaa{n_d}^2=
\averaa{n_d}(2-\averaa{n_d})-\Theta_S$, so for $\averaa{n_d}=1$ and $U=\infty$
we obtain $\mathrm{Var}\{n_d\}=0$.
The vanishing of charge fluctuations and the value of the spin--square 
$\averaa{\mathbf{S}_d^2}=3/4$ allows one to consider the localized spin 
$1/2$, a presence of which governs the ground--state properties at the 
strong--coupling limit. The density matrix (\ref{rhod}) takes thus the form
$\rho_d=(\ket{\ua}\bra{\ua}+\ket{\da}\bra{\da})/2$, which brought us to the 
value of the local entanglement $E_v=1$ at the 
particle--hole symmetric point. 
In contrast, for the noninteracting system all coefficients of the
density matrix (\ref{rhod}) are equal to $1/4$ and the local entanglement
reaches its maximal value $E_v=2$ (for $\averaa{n_d}=1$). 
The entanglement drop with the increasing coupling near the particle--hole 
symmetric point can therefore be explained as an effect of the formation of 
a~localized moment inside the dot.

The correspondence between inequalities (\ref{thlim}) and the limits defined 
by Eqs.\ (\ref{evzero}) and (\ref{evinfty}) become straightforward when 
expressing the coefficients of the density matrix (\ref{rhod}) as functions 
of $\averaa{n_d}$ and $\Theta_S$, what leads to the local entanglement
$$
  E_v = -\left(\frac{2-\averaa{n_d}-\Theta_S}{2}\right)\log_2
  \left(\frac{2-\averaa{n_d}-\Theta_S}{2}\right) 
$$
\begin{equation}
\label{evnth}
  - \left(\frac{\averaa{n_d}-\Theta_S}{2}\right)\log_2
    \left(\frac{\averaa{n_d}-\Theta_S}{2}\right)
    - \Theta_S\log_2\frac{\Theta_S}{2}.
\end{equation}
Eq.\ (\ref{evnth}) with the limits given by (\ref{thlim}) 
relates the entanglement between the
quantum dot and the leads to the local moment formation inside the dot.
It also express the local entanglement $E_v$ in terms of measurable quantities:
the dot occupation $\averaa{n_d}$ and the spin--square magnitude 
$\averaa{\mathbf{S}^2_d}$ contained in the parameter $\Theta_S$ (\ref{thes}). 
In contrast, the spin fluctuations are absent in Eq.\ (\ref{glutt}) for
the conductance $G$, which is fully determined by the dot filling 
$\averaa{n_d}$.
One can note Eq.\ (\ref{evnth}) is model--independent, providing we consider
the lattice system with one orbital per site. 
Thus, for the system with quantum phase transition, such as that considered by
Gu \emph{et al.}\ \cite{guet}, the nonanalytical behavior of the local 
entanglement $E_v$ is equivalent to the nonanalytical behavior of the 
spin--magnitude parameter $\Theta_S$.

\subsection{The fermionic concurrence}
We consider here the entanglement of two qubits, one associated with the 
electron localized on a quantum dot and other with the nearest one placed in 
a lead. 
The physical realization of individual qubits may, in principle, employ 
\emph{charge} or \emph{spin} degrees of freedom of the system in 
Fig.\ \ref{qdfig}.

The reduced density matrix (\ref{rhodef}) for the pair of electrons with equal 
spins (say $\sigma\equiv\sigma'=\ua$), one localized on a quantum dot and other
on a nearest lead atom ($i=d$, $j=j_{L(R)}$), can be written as \cite{deng}
\begin{equation}
\label{rhoij}
  \rho_{i\ua,j\ua} = 
  \left(\begin{array}{cccc}
    u_+^c & 0       & 0     & 0 \\
    0     & w_1^c   & z^c   & 0 \\
    0     & (z^*)^c & w_2^c & 0 \\
    0     & 0       & 0     & u_-^c \\
  \end{array}\right),
\end{equation}
where
$$
  u_+^c = \averaa{(1-n_{i\ua})(1-n_{j\ua})},\ \ \ \ 
  w_1^c = \averaa{n_{i\ua}(1-n_{j\ua})},
$$
\begin{equation}
\label{ijelem}
  w_2^c = \averaa{(1-n_{i\ua})n_{j\ua}},\ \ 
  z^c = \averaa{c^{\dagger}_{j\ua}c_{i\ua}},\ \ 
  u_-^c = \averaa{n_{i\ua}n_{j\ua}}.
\end{equation}
The upper index $c$ stands to denote, that the density matrix (\ref{rhoij})
refers to the \emph{charge} degrees of freedom, since the \emph{spin} direction
is arbitrarily chosen for both particles.

We use now the \emph{concurrence} $\mathcal{C}$ as a measure of the 
entanglement for such a two--qubit system. 
The closed--form expression, derived by Wootters \cite{woot}, reads
\begin{equation}
\label{conc:def}
  \mathcal{C} = \max\left\{0,\sqrt{\lambda_1}-\sqrt{\lambda_2}-
  \sqrt{\lambda_3}-\sqrt{\lambda_4}\right\}.
\end{equation}
The $\lambda_i$'s are the eigenvalues  of the matrix product 
$\rho\cdot (\sigma_i^y\otimes\sigma_j^y)\rho^* (\sigma_i^y\otimes\sigma_j^y)$,
and
$\lambda_1\geqslant\lambda_2\geqslant\lambda_3\geqslant\lambda_4$.
Since there exists a mo\-no\-to\-nous relation between the concurrence 
$\mathcal{C}$ and the entanglement of formation 
$E_f=-x\log_2x-(1-x)\log_2(1-x)$, where 
$x=1/2+\sqrt{1-\mathcal{C}^2}/2$ \cite{woot}, $\mathcal{C}$ is widely used  
instead of $E_f$ in the literature.

\begin{figure}[!t]
\begin{center}
\includegraphics[width=0.8\columnwidth]{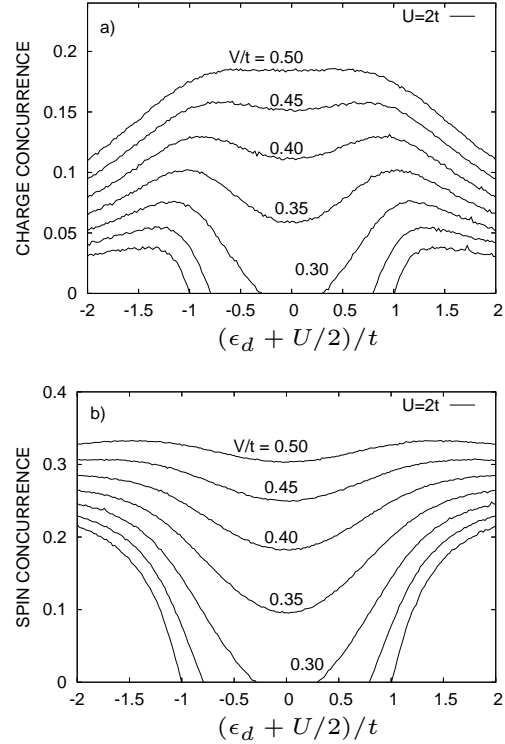}
\end{center}
\caption{
  The charge $(a)$ and spin $(b)$ pairwise concurrence for the system
  containing one qubit localized on quantum dot and other on the 
  nearest atom of the lead. 
}
\label{cocos}
\end{figure}

\begin{figure}[!t]
\begin{center}
\includegraphics[width=0.8\columnwidth]{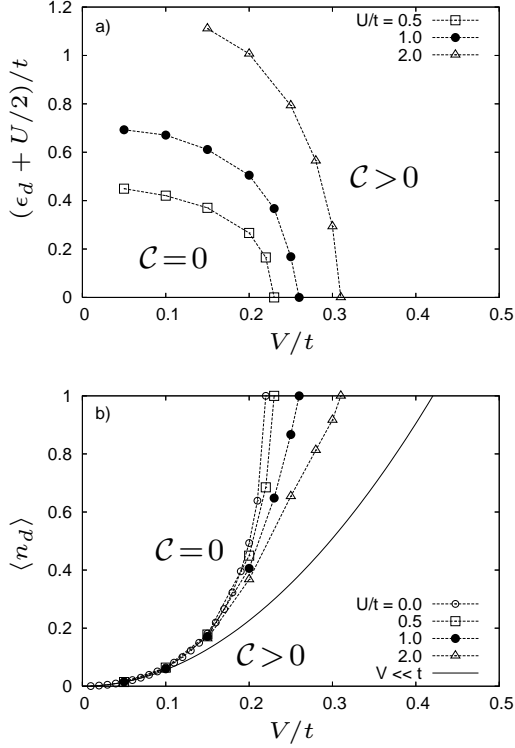}
\end{center}
\caption{
  The values of dot energy level $\epsilon_d$ $(a)$  and average occupation
  $\averaa{n_d}$ $(b)$ corresponding to zero concurrence, as a function of 
  hybridization $V$ and Coulomb interaction $U$ (specified for each dataset).
  The perturbative limit $V\ll t$ is also shown.
}
\label{gapco}
\end{figure}

For the density matrix $\rho_{i\ua,j\ua}$ (\ref{rhoij})
the corresponding concurrence can be calculated from Eq.\ (\ref{conc:def}) as
$$
  \mathcal{C}_{i\ua,j\ua}=2\max\left\{0,
  \abs{\averaa{c_{i\ua}^{\dagger}c_{j\ua}}}\right.
  \ \ \ \ \ \ \ \ \ \ \ \ \ \ \ \ \ \ \ \ \ \ \ \ \ \ \ \ \ \ \ \ 
  \ \ \ \ \ \ \ \ \ \ \ \ \ \ \ \ 
$$
\begin{equation}
\label{cconc}
  \ \ \ \ \ \ \ \ \ \ \ \ \ \ \ \ 
  \left.
  -\sqrt{\averaa{n_{i\ua}n_{j\ua}}\left(1-\averaa{n_{i\ua}}-\averaa{n_{j\ua}}
  +\averaa{n_{i\ua}n_{j\ua}}\right)}\right\}.
\end{equation}
Hereinafter, we call $\mathcal{C}_{i\ua,j\ua}$ a \textit{charge concurrence},
since it is related to the charge degrees of freedom.

Alternatively, one can consider the full two--site density matrix 
$\rho_{ij}=\mbox{Tr}_{ij}\ketaa{\Psi}\braaa{\Psi}$ 
(with $i=d$, and $j=j_{L(R)}$ again) and project out all the states except 
from these corresponding to $n_i=n_j=1$.
The resultant $4\times 4$ density matrix 
$\rho_{s_i,s_j} = \tilde{\rho}_{s_i,s_j}/\mbox{Tr}\tilde{\rho}_{s_i,s_j}$
describes the entanglement accessible by spin manipulation with 
a~particle--conservation constrain \cite{been}.
The matrix $\tilde{\rho}_{s_i,s_j}$ has a general structure of 
$\rho_{i\ua,j\ua}$ given by Eq.\ (\ref{rhoij}), with the elements 
(\ref{ijelem}) replaced by
$$
  u_+^s = \aver{n_{i\ua}(1-n_{i\da})n_{j\ua}(1-n_{j\da})},
$$
$$
  u_-^s = \aver{(1-n_{i\ua})n_{i\da}(1-n_{j\ua})n_{j\da}},
$$
\begin{equation}
\label{sselem}
  w_1^s = \aver{n_{i\ua}(1-n_{i\da})(1-n_{j\ua})n_{j\da}},
\end{equation}
$$
  w_2^s = \aver{(1-n_{i\ua})n_{i\da}n_{j\ua}(1-n_{j\da})},
$$
$$
  z^s = \aver{S_j^+S_i^-}=
  \averaa{c^{\dagger}_{j\ua}c_{j\da}c^{\dagger}_{j\da}c_{j\ua}}.
$$
The label $s$ indicates that we are working know with the \emph{spin} degrees 
of freedom, as charges of the $i$ and $j$ sites are chosen.
The concurrence, obtained by applying the definition (\ref{conc:def}) to
the density matrix $\rho_{s_i,s_j}$, is called a \textit{spin concurrence}.

The charge and spin concurrence is shown in Fig.\ \ref{cocos} as a function
of the dot energy level $\epsilon_d$. 
Again, we observe an excellent convergence of both the studied quantities for 
the system size of the order of $N\sim 1000$. 
Although the charge and spin concurrence are, in principle, two different 
physical quantities, they reach the limit $\mathcal{C}=0$ simultaneously for 
all the analyzed values of $V$ and $U$. 
Therefore, we can conclude that below the critical value
of the hybridization $V<V_c(U)$ and in the Kondo regime, the qubit localized on
the quantum dot is not entangled with other placed on the top of the lead
for \emph{neither} charge \emph{nor} spin degrees of freedom. 
This is because of large quantum fluctuations in each lead (modeled as a 
noninteracting Fermi gas), which therefore destroy the entanglement for a weak 
dot--lead coupling.

\begin{table}
\caption{
  The critical values of hybridization $V_c$ on the zero--concurrence 
  boundary for $\epsilon_d=-U/2$, the corresponding exchange coupling 
  $\rho(0)J_K=4V^2/\pi Ut$, and the Kondo temperature \cite{tksym}.
}
\begin{center}
\begin{tabular}{c|c|c|c}
\hline\hline
  $U/t$ & $V_c$ & $\rho(0)J_K$ & $T_K/t$ \\
    &  &  &  $[\mathrm{mK/eV}]$\\  \hline
  0.5 & 0.23 & 0.135 & 360 \\
  1.0 & 0.26 & 0.086 & 8.6 \\
  2.0 & 0.31 & 0.061 & 0.13 \\ 
\hline\hline
\end{tabular}
\end{center}
\end{table}

The values of $\epsilon_d$ and $\averaa{n_d}$, corresponding to $\mathcal{C}=0$
are depicted in Fig.\ \ref{gapco}. 
To complement the analysis we also provide, in Table~1, the values of the 
symmetric Kondo temperature \cite{tksym}, corresponding to the critical 
hybridization $V_c(U)$, at which the concurrence vanish in the ground state. 
Although the numerical results presented in Fig.\ \ref{gapco} refers to
$T=0$, it is clear that in the finite--$T$ situation the conductance 
$G\approx G_0$ at the particle--hole symmetric point if $T\ll T_K$, 
and that the concurrence $\mathcal{C}$ decrease with $T$ due to dephasing. 
Therefore we can conclude, that the coexistence of zero concurrence and 
maximal conductance can be observed for $U\leqslant t$ if the relative 
temperature $T/t$ is of the order of a few $\mathrm{mK/eV}$.
Another interesting feature of these results is
the universal (interaction independent) behavior of the maximal dot 
filling $\averaa{n_d}=\averaa{n_d}_\mathrm{max}$, for which the entanglement 
$\mathcal{C}\geqslant 0$, with $V/t\rightarrow 0$ 
(\emph{cf.} Fig.\ \ref{gapco}b). 
This observation can be rationalized by using Eq.\ (\ref{cconc}) and putting 
$\mathcal{C}_{i\ua,j\ua}=0$. Then, in the perturbative limit $V\ll t$, we 
obtain 
$$
  \averaa{n_d}_\mathrm{max}\approx 2^{3/2}\absa{\averaa{c_i^{\dagger}c_j}}
  \approx 2^{5/2}(V/t)^2, 
$$
up to the quadratic terms. The agreement with the numerical 
data is perfect for $V/t\leqslant 0.1$.


\section{Summary}
We analyzed the local entanglement between the quantum dot and the leads
as a function of the dot energy level $\epsilon_d$, the dot--lead 
hybridization $V$ and the intra--dot Coulomb repulsion $U$. 
The measure of this
entanglement, the von Neumann entropy $E_v$, evolves gradually from
the \emph{weak--coupling} limit, in which the maximal $E_v$ 
match the maximal quantum value of the conductance $G=G_0=2e^2/h$, 
to the \emph{strong--coupling} situation, where maximal $G$ corresponds
to the local minimum of $E_v$.
This behavior was explained in terms of local moment formation inside the dot,
which took place when the charge transport is dominated by the Kondo effect.

Finally,
we defined the pairwise concurrence, measuring the entanglement between
a pair of qubits: one localized on the dot and other on the nearest atom of 
the lead, for \emph{charge} and \emph{spin} degrees of freedom separately. 
Both quantities vanish simultaneously in the
Kondo--resonance range, where the weakly--entangled system
show the maximal conductance. 
We predict the latter to be observable at the temperature range of
$T/W\sim 1\ \mathrm{mK/eV}$ (where $W=4t$ is the lead bandwidth), 
which seems to be accessible within the present nanoscale experimental 
techniques. 
The universal dependence of the maximal
dot filling, above which the concurrence vanish, 
$\aver{n_d}_\mathrm{max}\approx 2^{5/2}(V/t)^2$ for $V/t \ll 1$, was also 
identified.

\section*{Acknowledgment}
I am grateful to Prof.\ C.\ W.\ J.\ Beenakker for many discussions and 
comments.
Remarks by Profs.\ M.\ A.\ Martin--Delgado, A.\ M.\ Ole\'{s}, J.\ Spa{\l}ek, 
and  Drs.\ B.\ Trauzettel, D.\ Sanchez are appreciated.
The work was supported by the Polish Science Foundation (FNP) Foreign Postdoc 
Fellowship, and by Polish Mi\-nis\-try of Education and Science, 
Grant No.\ 1 P03B 001 29.



\vspace{2em}
\begin{thebibliography}{99}
\bibitem{epr}
A.\ Einstein, B.\ Podolski, and N.\ Rosen, 
\textit{Phys.\ Rev.} \textbf{47}, 777 (1935).
\bibitem{qinfo}
See review by C.\ H.\ Bennet and D.\ P.\ Divincenzo, 
\textit{Nature} \textbf{404}, 247 (2000);
M.\ A.\ Nielsen and I.~L.~Chuang, 
\textit{Quantum Computation and Quantum Information}
(Cambridge, 2000).
\bibitem{qptra}
S.\ Sachdev, \textit{Quantum Phase Transitions} 
(Cambridge University Press, Cambridge, 2000).
\bibitem{spinc}
A.\ Osterloh \emph{et al.}, 
\textit{Nature} \textbf{416}, 608 (2002);
T.\ J.\ Osborne and M.\ A.\ Nielsen, 
\textit{Phys.\ Rev.\ A} \textbf{66}, 032110 (2002);
V.\ Subrahmanyam, 
\emph{ibid.} \textbf{69}, 022311 (2004).
\bibitem{delga}
F.\ Verstraete, M.\ A.\ Martin--Delgado, J.\ I.\ Cirac,
\textit{Phys.\ Rev.\ Lett.} \textbf{92}, 087201 (2004);
M.\ Popp \emph{et al.}, \texttt{quant-ph/0411123}, unpublished. 
\bibitem{oles}
A.\ M.\ Ole\'{s} \emph{et al.}, 
\textit{Phys.\ Rev.\ Lett.} \textbf{96}, 147205 (2006). 
\bibitem{webza}
J.\ van Wezel, J.\ van den Brink, J.\ Zaanen, 
\textit{Phys.\ Rev.\ Lett.} \textbf{94}, 230401 (2005). 
\bibitem{schli}
J.\ Schliemann, D.\ Loss, A.\ H.\ MacDonald,
\textit{Phys.\ Rev.\ B} \textbf{63}, 085311 (2001);
J.\ Schliemann \emph{et al.}, 
\textit{Phys.\ Rev.\ A} \textbf{64}, 022303 (2001).
\bibitem{zanar}
P.\ Zanardi, 
\textit{Phys.\ Rev.\ A} \textbf{65}, 042101 (2002);
P.\ Zanardi, X.\ Wang,
\textit{J.\ Phys. A} \textbf{35}, 7947 (2002).
\bibitem{guet}
S.--J.\ Gu \emph{et al.}, 
\textit{Phys.\ Rev.\ Lett.} \textbf{93}, 086402 (2004).
\bibitem{coher}
A.\ O.\ Caldeira and A.\ J.\ Leggett, 
\textit{Phys.\ Rev.\ Lett.} \textbf{46}, 211 (1981);
I.\ L.\ Chuang \emph{et al.}, 
\textit{Science} \textbf{270}, 1633 (1995).
\bibitem{choi}
M.--S.\ Choi, R.\ L\'{o}pez, R.\ Aguado, 
\textit{Phys.\ Rev.\ Lett.} \textbf{95}, 067204 (2005);
R.\ L\'{o}pez \emph{et al.},
\textit{Phys.\ Rev.\ B} \textbf{71}, 115312 (2005).
\bibitem{woot}
W.\ K.\ Wootters, 
\textit{Phys.\ Rev.\ Lett.} \textbf{80}, 2245 (1998);
S.\ Hill, W.\ K.\ Wootters, 
\textit{Phys.\ Rev.\ Lett.} \textbf{78}, 5022 (1997).
\bibitem{bean}
P.\ B.\ Wiegman, A.\ M.\ Tsvelick, 
\textit{Pis'ma ZETF} \textbf{35}, 100 (1982);
\textit{J.\ Phys.\ C} \textbf{16}, 2281 (1983).
\bibitem{hofs}
W.\ Hofstetter, J.\ K\"{o}nig, and H.\ Schoeller, 
\textit{Phys.\ Rev.\ Lett.} \textbf{87} 156803 (2001).
\bibitem{meir}
Y.\ Meir, N.\ S.\ Wingreen, 
\textit{Phys.\ Rev.\ Lett.} \textbf{68}, 2512 (1992);
A.--P.\ Jauho \emph{et al.}, 
\textit{Phys.\ Rev.\ B} \textbf{50}, 5528 (1994).
\bibitem{corn}
P.\ S.\ Cornaglia \emph{et al.}, 
\textit{Phys.\ Rev.\ Lett.}\ \textbf{93}, 147201 (2004).
\bibitem{bulk}
P.\ Stefa\'{n}ski, A.\ Tagliacozzo, B.\ Bu{\l}ka,
\textit{Phys.\ Rev.\ Lett.}\ \textbf{93}, 186805 (2004);
B.\ Bu{\l}ka, P.\ Stefa\'{n}ski,
\emph{ibid.}\ \textbf{86}, 5128 (2001).
\bibitem{rera}
T.\ Rejec, A.\ Ram\v{s}ak, 
\textit{Phys.\ Rev.\ B} \textbf{68}, 035342 (2003).
\bibitem{reram}
T.\ Rejec, A.\ Ram\v{s}ak, 
\textit{Phys.\ Rev.\ B} \textbf{68}, 033306 (2003).
\bibitem{ryspa}
A.\ Rycerz, J.\ Spa{\l}ek, \texttt{cond-mat/0604237}, to be published.
\bibitem{edabi}
A.\ Rycerz, J.\ Spa{\l}ek, 
\textit{Eur.\ Phys.\ J.\ B} \textbf{40}, 153 (2004);
\textit{Phys.\ Rev.\ B} \textbf{63}, 073101 (2001); 
\textit{ibid.}\ \textbf{65}, 035110 (2002).
\bibitem{pabc}
E.\ H.\ Lieb, 
\textit{Phys.\ Rev.\ Lett.} \textbf{73}, 2158 (1994);
F.\ Nakano, 
\textit{J.\ Phys. A} \textbf{33}, 5429 (2000);
\emph{ibid.} \textbf{37}, 3979 (2004).
For a discussion of boundary condition role in even/odd effect
for correlated nanosystems see: 
A.\ Rycerz, J.\ Spa{\l}ek, 
\textit{phys.\ stat.\ sol.} (b) \textbf{243}, 183 (2006).
\bibitem{half}
The notion of the \emph{half--filling} refers to the entire system composed
of a quantum dot and leads, as presented in Fig.\ \ref{qdfig}. 
The average dot occupation $\averaa{n_d}$ is determined by the energy level
$\epsilon_d$, and reaches $1$ at the particle--hole symmetric point 
$\epsilon_d=-U/2$.
\bibitem{kondo}
J.\ Kondo, 
\textit{Prog.\ Theor.\ Phys.}\ \textbf{28}, 846 (1962);
J.\ R.\ Schrieffer, P.\ A.\ Wolf,
\textit{Phys.\ Rev.}\ \textbf{149}, 491 (1966).
\bibitem{tksym}
The symmetric Kondo temperature for $\epsilon_d=-U/2$ \cite{bean} is given by
$T_K=\sqrt{2U\Gamma/\pi^2}\exp(-\pi U/8\Gamma)$, with the impurity level
width $\Gamma=\pi\rho(\epsilon_F)V^2=V^2/\sqrt{4t^2-\epsilon_F}$, where the
second equality refers to the tight--binding electrodes shown 
in~Fig.\ \ref{qdfig}.
\bibitem{hews}
A.\ C.\ Hewson, 
\textit{The Kondo Problem to Heavy Fermions}
(Cambridge University Press, 1997).
\bibitem{spal}
J.\ Spa{\l}ek, 
\textit{J.\ Sol.\ St.\ Chem.}\ \textbf{88}, 70 (1990);
J.\ Spa{\l}ek, A.\ M.\ Ole\'{s}, J.\ M.\ Honig, 
\textit{Phys.\ Rev.\ B} \textbf{28}, 6802 (1983).
\bibitem{deng}
For an application to the Hubbard dimer,
see: S.--S.\ Deng, S.--J.\ Gu, H.--Q.\ Lin,
\textit{Chin.\ Phys.\ Lett.}\ \textbf{22}, 804 (2005).
The ge\-ne\-ra\-li\-za\-tion for any two neighboring sites of 
one--dimensional system with one orbital per site is straightforward.
\bibitem{been}
See, e.g.\ C.\ W.\ J.\ Beenakker, \texttt{cond-mat/0508488}, to be published
in \textit{Quantum Computers, Algorithms and Chaos}, Int.\ School
of Phys.\ ``Enrico Fermi'', vol.\ 162, and references therein. 
\end{thebibliography}
\end{document}